\documentclass{article}
\usepackage[utf8]{inputenc}
\usepackage{spconf,amsmath,graphicx}
\usepackage{multirow,multicol,url}
\usepackage{todonotes}
\graphicspath{{./}}

\title{Sound event detection and separation: a benchmark on DESED synthetic soundscapes}
\name{
    \begin{tabular}{@{}c@{}}
    Nicolas Turpault$^1$\thanks{Part of this work was made with the support of the French National Research Agency, in the framework of the  project LEAUDS “Learning to under-stand audio scenes” (ANR-18-CE23-0020) and the French region Grand-Est. High Performance Computing resources were partially provided by the EXPLOR centre hosted by the University de Lorraine.},
    Romain Serizel$^1$,
    Scott Wisdom$^2$,
    Hakan Erdogan$^2$,\\
    John R. Hershey$^2$,
          Eduardo Fonseca$^3$,
    Prem Seetharaman$^4$,
    Justin Salamon$^5$
    \end{tabular}
    }
\address{$^1$ Universit{\'e} de Lorraine, CNRS, Inria, Loria, F-54000 Nancy, France \\
        $^2$ Google Research, AI Perception, Cambridge, United States\\
        $^3$ Music Technology Group, Universitat Pompeu Fabra, Barcelona\\
        $^4$Descript, Inc., San Francisco, United States\\
        $^5$ Adobe Research, San Francisco, United States
 }

\date{October 2020}
\ninept
\begin{document}

\maketitle

\begin{sloppy}

\begin{abstract}
We propose a benchmark of state-of-the-art sound event detection systems (SED). We designed synthetic evaluation sets to focus on specific sound event detection challenges. We analyze the performance of the submissions to DCASE 2021 task 4 depending on time related modifications (time position of an event and length of clips) and we study the impact of non-target sound events and reverberation. We show that the localization in time of sound events is still a problem for SED systems. We also show that reverberation and non-target sound events are severely degrading the performance of the SED systems. In the latter case, sound separation seems like a promising solution.

\end{abstract}


\section{Introduction}
Sound event detection (SED) is a task from ambient sound analysis that consists in detecting what did happen in an audio recording but also when it did happen~\cite{virtanen2018computational}. SED can have numerous applications from assisted living to security or urban planning~\cite{Bello:SONYC:CACM:18,radhakrishnan2005audio,serizel2016, zigel2009method}. To be useful for these applications SED algorithms have to be efficient when applied in real-world, complex scenarios.

SED in real environments includes several challenges among which detecting accurately the sound events time boundaries (onset and offset) in scenarios that often include several overlapping sound events (target or not) and possibly background noise. One key aspect in solving this problem is to have training set with strong annotations (with timestamps) available. This is rarely the case because this type of annotations is time consuming to obtain and is known to be prone to annotation errors and disagreement between annotators (mainly because of the ambiguity in the perception of the onsets and offsets).

In 2018, in the task 4 of the international challenge on detection and classification of acoustic scenes and events (DCASE) we proposed to train a SED system from a dataset composed of recorded soundscapes partly labeled with weak annotations (without timestamps)~\cite{serizel2018_DCASE}. In 2019 we extended the training dataset with a subset composed of strongly annotated synthetic soundscapes (that are quite cheap to generate)~\cite{Turpault2019_DCASE}. In 2020 we proposed to use sound separation (SSep) as a pre-processing to SED in order to solve the problem of detecting overlapping sound events~\cite{turpault:hal-02891700}.

One problem is that the evaluation on complex recorded soundscapes does not allow to disentangle the several challenges faced in SED in real environments. During DCASE 2019 task 4, capitalizing on the possibility to have a full control on the properties of the soundscapes generated with Scaper~\cite{salamon2017scaper}, we benchmarked SED submissions on soundscapes designed to investigate several SED challenges such as foreground event to background ratio or time localization of the sound event within a clip challenges~\cite{Serizel2020_ICASSP}. During DCASE 2020 task 4 we proposed a new set of synthetic soundscapes designed to investigate other SED challenges.

This paper presents a benchmark of state of the art SED systems submitted to DCASE 2020 task 4 on several SED challenges. We analyze the performance of this year submissions depending on the time localization of the sound events within the clip, the sound events duration. We also compare the performance of the submitted systems on 10 second audio clips (as in the official evaluation set) and on 60 second audio clips. Finally, since one of the novelty of DCASE 2020 task 4 was to propose to use SSep as a pre-processing to SED, we analyze the robustness of the submissions performance towards noise and reverberation for both SED systems with or without SSep.

\section{Datasets and task setup}

\subsection{Task setup and evaluation metrics}

In DCASE 2020 task 4, systems are expected to produce strongly-labeled outputs (i.e.~detect sound events with a start time, end time, and sound class label), but are provided with weakly labeled data (i.e.~sound recordings with only the presence/absence of a sound event included in the labels without any timing information) for training. Multiple events can be present in each audio recording, including overlapping target sound events and potentially non-target sound events. Previous studies have shown that the presence of additional sound events can drastically decrease the SED performance~\cite{Serizel2020_ICASSP}.

In this paper, we evaluate SED submissions according to an event-based F-score with a 200~ms collar on the onsets and a collar on the offsets that is the greatest between 200~ms and 20\% of the sound event's length. The overall F-score is the unweighted average of the class-wise F-scores. F-scores are computed on a single operating point (fixed decision thresholds) using the sed\_eval library~\cite{mesaros_metrics_2016}. For a detailed analysis of the submissions performance depending on the evaluation metric we refer the reader to Ferroni et al.~\cite{ferroni_2021}.

\subsection{Datasets}

\subsubsection{DESED dataset}

The dataset used for the SED experiments is DESED\footnote{\url{https://project.inria.fr/desed/}}, a dataset for SED in domestic environments composed of 10-sec audio clips that are recorded or synthesized~\cite{Serizel2020_ICASSP,turpault:hal-02160855}. The recorded soundscapes are taken from AudioSet~\cite{Gemmeke2017audioset}. The synthetic soundscapes are generated using Scaper~\cite{salamon2017scaper}. The foreground events are obtained from FSD50k~\cite{font2013freesound,fonseca2020fsd50k}. The background textures are obtained from the SINS dataset~\cite{Dekkers2017} and TUT scenes 2016 development dataset~\cite{mesaros2016tut}.

In this benchmark we focus mainly on the evaluation sets composed of synthetic soundscapes. The performance of the submissions on the official evaluation set (composed of recorded soundscapes) is still reminded in Table~\ref{tab:rank} as a reference point.

\subsubsection{FUSS dataset}

The evaluation set of the Free Universal Sound Separation (FUSS)\footnote{\url{https://github.com/google-research/sound-separation/tree/master/datasets/fuss}} dataset \cite{wisdom2020fuss} is intended for experimenting with universal sound separation \cite{kavalerov2019universal}, and is used to generate the soundscapes used to investigate the robustness of the submission to additive noise and reverberation.
Audio data is sourced from \url{freesound.org}. Using labels from FSD50k \cite{fonseca2020fsd50k}, gathered through the Freesound Annotator \cite{Fonseca2017freesound}, these source files have been screened such that they likely only contain a single type of sound. Labels are not provided for these source files, and thus the goal is to separate sources without using class information.
To create reverberated mixtures, 10 second clips of sources are convolved with simulated room impulse responses (RIR). Each 10 second mixture contains between 1 to 4 sources (target sound event or not). Source files longer than 10 seconds are considered "background" sources. Every mixture contains one background source, which is active for the entire duration.

\subsection{Synthetic evaluation datasets}
The SED aspects that we are aiming to investigate here are the challenges related to timing (clips duration, sound events duration and sound events localization in time), to overlapping target/non-target sound events and reverberation faced in real scenarios. For this we designed five evaluation sets that are composed on synthetic soundscapes designed specifically to target these challenges. These additional evaluation datasets were proposed to DCASE 2020 task 4 participants together with the official evaluation set in order to be able to benchmark state-of-the-art systems on these particular challenges.

\subsubsection{Reference synthetic soundscapes evaluation set}
A reference subset comprised of 828 soundscapes is generated with Scaper scripts that are designed such that the distribution of sound events per class, the number of sound events per clip (depending on the class) and the sound event class co-occurrence are similar to that of the validation set which is composed of real recordings. The foreground event to background signal-to-noise ratio (FBSNR) parameter was uniformly drawn between 6~dB and 30~dB. Unless mentioned otherwise, the Sox based reverb in Scaper is deactivated in all the subsets. This subset is thereafter refereed to as \textbf{ref}.

\subsubsection{60~s clips}
A subset of 152 soundscapes is generated with a similar sound events number per clips and per class as in \textbf{ref} and with the same FBSNR range. The clip duration in this subset is 60~s which means that the sound event density relative to time is scaled by a factor 6. The motivation for this subset is that in real scenarios, SED systems are usually required to operate on sound segments that are longer than 10~s. Additionally, when operating in real life scenarios, SED systems would probably sometimes face scenarios where the sound events density is much lower than the sound event density faced in Youtube video that are generally recorded because something is actually happening. The subset is thereafter refereed to as \textbf{60s}. Note that this is the only subset with a clip duration which is not 10~s.

\subsubsection{Varying onset time}
\label{sub:seg}
A subset of 1000 soundscapes is generated with uniform sound event onset distribution and only one event per soundscape. The parameters are set such that the FBSNR is between 6~dB and 30~dB. Three variants of this subset are generated with the same isolated events, only shifted in time. In the first version, all sound events have an onset located between 250~ms and 750~ms, in the second version the sound event onsets are located between 5.25~s and 5.75~s and in the last version the sound event onsets are located between 9.25~s and 9.75~s. In the remainder of the paper, these subsets will be referred to as \textbf{500ms}, \textbf{5500ms} and \textbf{9500ms}, respectively. This subset is designed to study the sensibility of the SED segmentation to the sound event localization in time. In particular, we wanted to control if SED systems were learning a bias in term of time localization depending on the event length (e.g., long sound events would most often start at the beginning of the sound clip). Note that we already conducted experiments on a similar susbset in DCASE 2019 task 4~\cite{Serizel2020_ICASSP}. The purpose here is to analyze if systems have improved on this particular aspect.

\subsubsection{One event per file}
A subset of 1000 soundscapes was generated with a single event per clip. The FBSNR is drawn randomly between  6~dB and 30~dB. The time localization of the sound event onset is drawn randomly. The clip distribution per sound event class is uniform (\textit{i.e.}, there is 100 clips for each of the 10 sound event classes). The purpose of this subset is to analyze the impact of the sound events duration on the SED performance. It has been shown previously that some systems can have a tendency to perform better on long sound events than on short sound events~\cite{serizel2019} but this was analyzed in recorded clips where long sound events are more likely to be present alone (at least during a portion of the sound event) than short sound events. Another study confirmed that the signal-to-noise ratio between sound events does have a serious impact on the SED performance (both for short and long sound events)~\cite{Serizel2020_ICASSP}. Therefore, we propose this subset with only one sound event per clips that would allow to focus on events duration only and leave aside the overlapping events effect. This subset is thereafter refereed to as \textbf{single}.

\subsubsection{Modify conditions}

Eight additional versions of \textbf{ref} are generated to include non-target sound events, reverberation or a combination of both. The non target sound events are randomly selected from the FUSS dataset~\cite{wisdom2020fuss}. The target sound event to non-target sound event SNR (TNTSNR) can be set to 15~dB or 0~dB. The subsets without non target sound events, with TNTSNR at 15~dB and at 0~dB will be respectively refereed to as \textbf{TNTSNR\_inf}, \textbf{TNTSNR\_15} and \textbf{TNTSNR\_0}. Reverberation is applied using RIR from the FUSS dataset~\cite{wisdom2020fuss}. Each soundscape is reverberated with a different room from the FUSS dataset (each sound event is convolved with a RIR corresponding to a different location in the room). The reverberation can be applied either using the full RIR or applying a RIR truncated to 200~ms after the direct path. The subsets without reverberation, with reverberation from truncated RIR and with full RIR will be refereed to as \textbf{no\_reverb}, \textbf{short\_reverb} and \textbf{long\_reverb}, respectively. The 8 subsets correspond to the combinations of the TNTSNR and the reverberation conditions. The subset with \textbf{TNTSNR\_inf} and \textbf{no\_reverb} is \textbf{ref}.

\section{Analysis on time related variations}
\subsection{60~s clips}

\begin{table}[t]
\centering
\begin{tabular}{l||rrr|r}

\multirow{2}{*}{Submission} &   \multicolumn{3}{c|}{F-score}&Difference
\\
&\textbf{2020 Eval}  &  \textbf{ref} &  \textbf{60s} & \textbf{60s-ref}\\
\hline \hline

         \textbf{Miyazaki~\cite{Miyazaki2020}} &       51.1 &                               56.5 &                    2.9 &            -53.6
         \\
             \textbf{Hao~\cite{Hao2020}} &       47.8 &                               38.4 &                   53.0 &             14.7
             \\
          \textbf{Ebbers~\cite{Ebbers2020}} &       47.2 &                               54.4 &                   53.0 &             -1.5
          \\
            \textbf{Koh~\cite{Koh2020}} &       46.6 &                               51.4 &                    3.3 &            -48.1
            \\
          \textbf{Yao~\cite{Yao2020}} &       46.4 &                               52.9 &                    2.7 &            -50.2        
          \\
              \textbf{CTK~\cite{Chan2020}} &       46.3 &                               50.9 &                   39.9 &            -11.1
              \\
         \textbf{Liu~\cite{Liu2020}} &       45.2 &                               45.3 &                   41.7 &             -3.6    
         \\
         \textbf{Zhenwei~\cite{HouZ2020}} &       45.1 &                               36.4 &                    0.1 &            -36.3  
         \\
   \textbf{Huang~\cite{Huang2020}} &       44.7 &                               36.2 &                   35.8 &             -0.4         
   \\
   \textbf{Cornell~\cite{Cornell2020}} &       44.4 &                               47.8 &                    3.4 &            -44.4        
   \\
\hline
            Baseline~\cite{turpault:hal-02891665} &       36.5 &                               46.3 &                    3.0 &            -43.3
            \\
\end{tabular}
\label{tab:rank}
\caption{F-score performance of DCASE 2020 task 4 submissions on the official evaluation set and synthetic evaluation sets.}
\end{table}

In Table~\ref{tab:rank} we present the F-score performance obtained by the 10 teams that were at the highest ranks during the DCASE 2020 task 4 and for the baseline. We present F-score for the official evaluation set (\textbf{2020 Eval}) and on the synthetic sets \textbf{ref} and \textbf{60s}. We also present the difference between the performance obtained on \textbf{ref} to the performance obtained on \textbf{60s}.

For most of the systems the performance is severely degraded when evaluating on \textbf{60s} instead of \textbf{ref}. This is not the case for three systems that maintain the performance~\cite{Ebbers2020,Liu2020,Huang2020} and one system that even improves its performance~\cite{Hao2020}. The difference of behavior compared to other systems could be related to the specific attention some of these participants spent on adjusting the decision thresholds~\cite{Ebbers2020,Liu2020} that can be crucial in avoiding false negatives or false positives in particular when the sound event density decreases.

For systems on which we observe degraded performance on \textbf{60s}, our first assumption was that the performance difference was caused by the change in sound events distribution. This could have caused systems trained on 10~s clips to predict more sound events on \textbf{60s} than the actual distribution causing false positives. This hypothesis is actually refuted when looking at the precision and recall performance of the systems (Figure~\ref{fig:long_clips}). The low recall indicates a high number of false negatives and not false positives. This could be due to the system lack of ability to segment sound event on longer clips or on the bias introduced by event-based (collar-based) metric on long sound events performance evaluation~\cite{ferroni_2021}. This hypothesis should probably be confirmed with another metric like polyphonic sound detection score~\cite{PSDS:2020} but this is out of the scope of the paper.

\begin{figure}[h]
	\includegraphics[width=\linewidth]{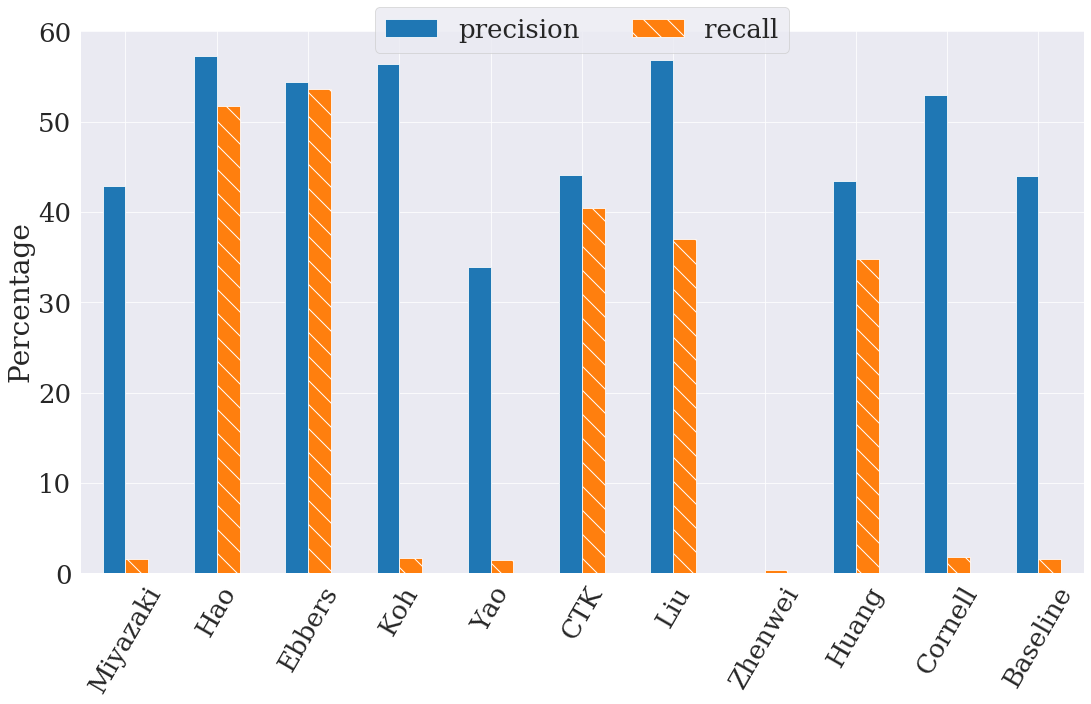}
	\caption{Submissions performance on (\textbf{60s}): Precision and recall}
	\label{fig:long_clips}
\end{figure}


\subsection{Impact of the time localization of events}
This experiment is a follow-up to the experiment proposed on DCASE 2019 task 4 submissions~\cite{Serizel2020_ICASSP}. The time localization of the sound event within the clip has only a minor impact on the detection of short events so we focus on long sound events here. Figure~\ref{fig:onsets} shows the detection performance for long sound events depending on where the sound event is located within the clip. The performance degrades when the event is located towards the end of the clip. As the onset distribution in the training set is uniform in the clip~\cite{Serizel2020_ICASSP} this is most probably due to post-processing that are based on too long windows for long sound events.

\begin{figure}[h]
	\includegraphics[width=\linewidth]{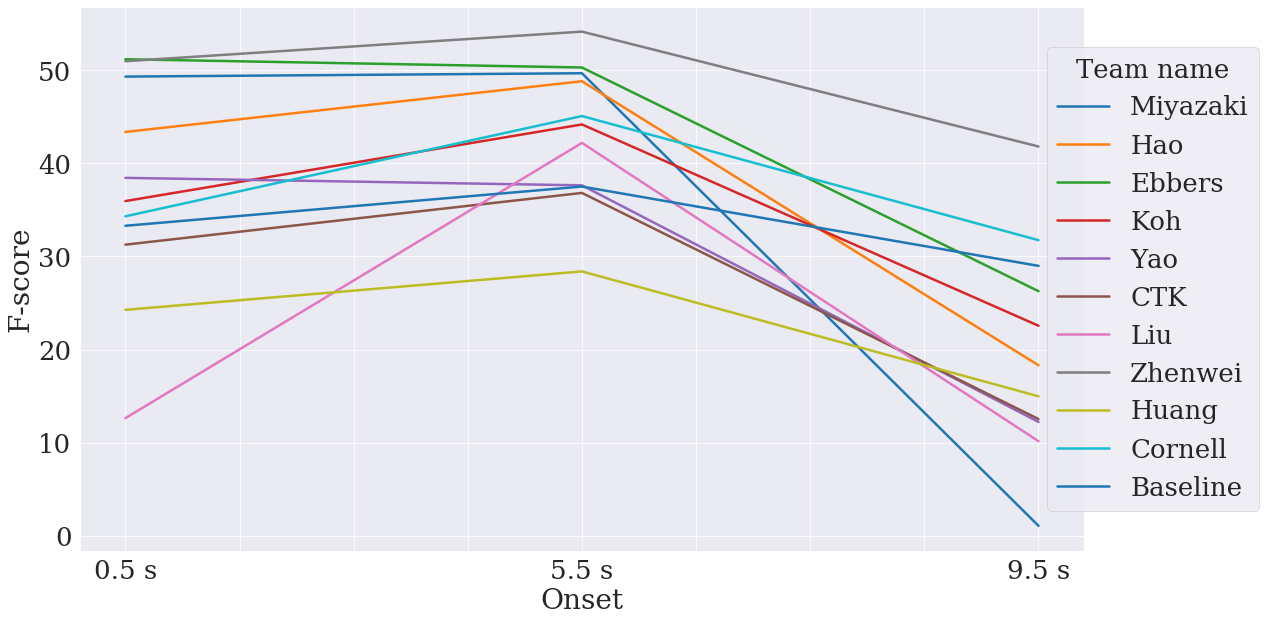}
	\caption{F-score depending on the event localization in time.}
	\label{fig:onsets}
\end{figure}

\subsection{One event per file}
The F-score of the submissions on \textbf{single} is presented in Figure~\ref{fig:one_event}. In general SED systems tend to perform better on short sound events than on long sound events. The observation that SED systems performed better on long sound events when performing SED on recorded clips~\cite{serizel2019} was then probably due to the impact of event polyphony and not the sound event duration. The poor performance on long sound events could also be emphasized by the biased introduced by collar based metrics~\cite{ferroni_2021}.

\begin{figure}[h]
	\includegraphics[width=\linewidth]{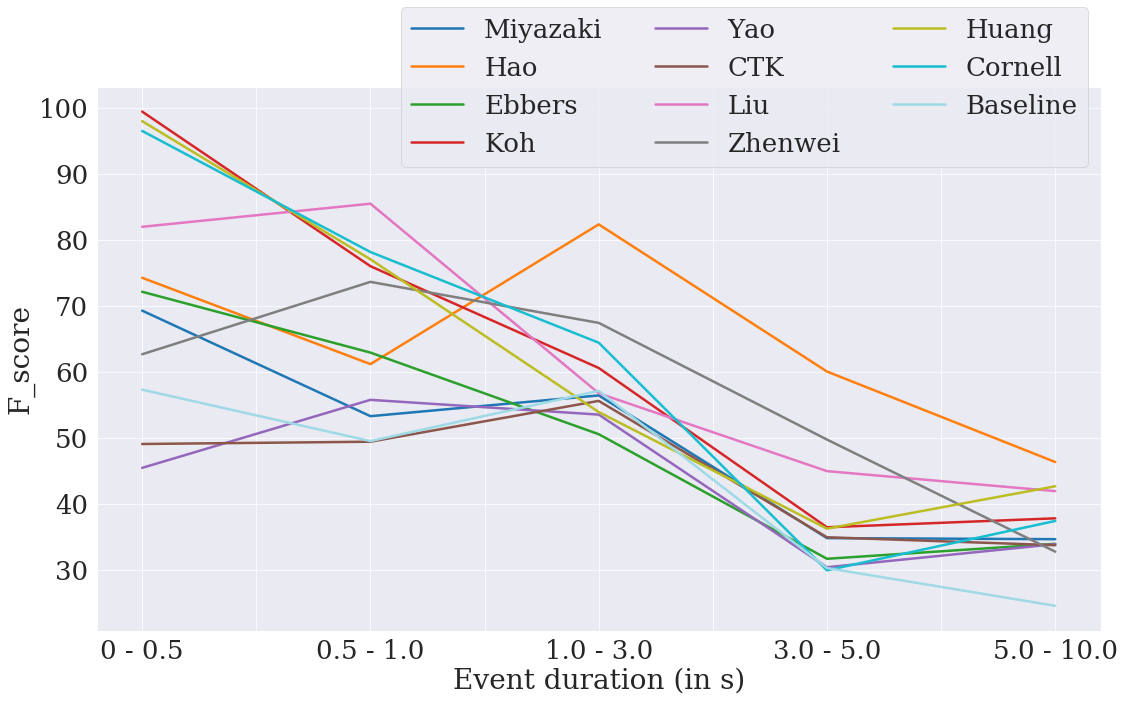}
	\caption{F-score depending on the sound events duration}
	\label{fig:one_event}
\end{figure}


\section{Robustness to additive noise and reverberation}
In Figures~\ref{fig:snr} and~\ref{fig:reverb} we present systems performance depending on the TNTSNR and the reverberation, respectively. The general trend is the same for all systems: the performance decreases when the TNTSNR decreases and the performance also decreases when the soundscapes are reverberated (the degradation is rather similar with short and long reverberation). The performance degradation when introducing reverberation is about 15\% in terms of F-score in average. This can be problematic when considering SED on real recordings where the reverberation can change highly from one recording to another. The impact of reverberation on SED systems was not studied until now but these results open the way to experiments on multi-condition training or on using dereverberation with SED.

\begin{figure}[h]
	\includegraphics[width=\linewidth]{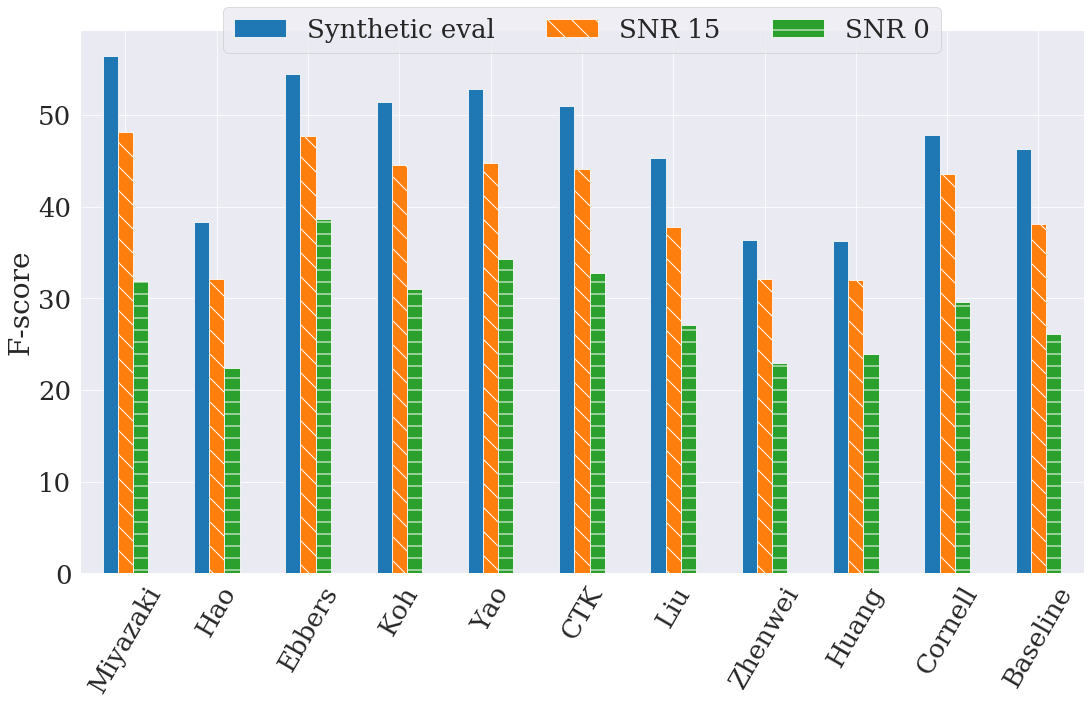}
	\caption{F-score depending on the TNTSNR (without reverberation)}
	\label{fig:snr}
\end{figure}

\begin{figure}[h]
	\includegraphics[width=\linewidth]{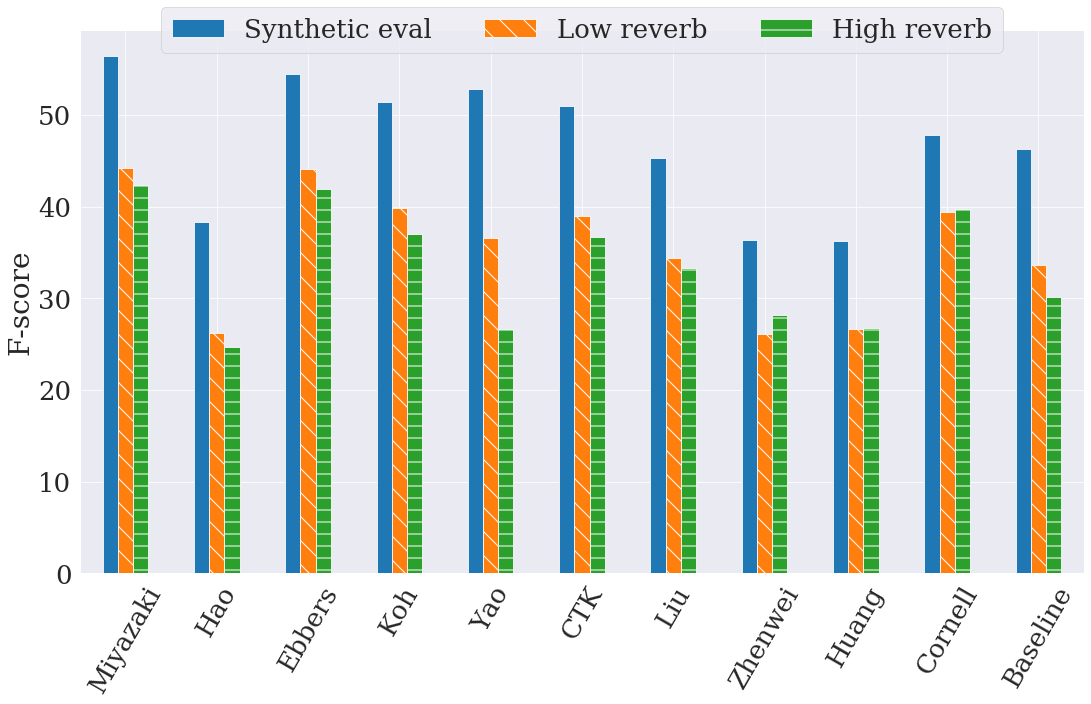}
	\caption{F-score depending on the reverberation (without non-target events)}
	\label{fig:reverb}
\end{figure}

%
%

As the general trends for SED performance depending on the TNTSR is rather similar across systems, we propose to analyze systems performance in average depending on whether the submission includes a SSep pre-processing or not (Table~\ref{tab:noSSepvsSSep}). SSep fails to improve the performance when no non-target sound events are present in the recording (\textbf{TNTSNR\_inf}). This is true for the baseline. Regarding the submissions, the SED systems used with or without SSep are different so the performance cannot be compared directly. We would need several submissions that used the same SED without and without SSep to analyze the absolute impact of SSep on SED (which we did not have this year). When comparing the relative robustness to non-target sound events the degradation is about 19\% in terms of F-score from \textbf{TNTSNR\_inf} to \textbf{TNTSNR\_0} for submissions wihtout SSep while it is about 12.5\% for submissions with SSep. This is also true to a lesser extent for the baselines.

\begin{table}[h]
\centering
\begin{tabular}{l|l||r|r|r}
& &\multicolumn{3}{c}{TNTSNR}\\
      & SSep      & 0 & 15 & inf\\ \hline\hline
\multirow{2}{*}{Baseline} & No & 25.68 & 38.65  & 48.16   \\
&Yes & 26.13 & 38.15  & 46.27   \\ \hline\hline
Submissions&No  & 29.66 & 41.65  & 48.44  \\
(average)&Yes  & 24.23 & 32.69  & 36.78
\end{tabular}
\caption{F-score average performance for the submissions and the baseline with/without SSep}
\label{tab:noSSepvsSSep}
\end{table}
\section{Conclusions}
\label{sec:conclusions}

In this paper we performed a benchmark of state-of-the-art SED systems (the submissions to DCASE 2020 task 4) on evaluation sets composed of synthetic soundscapes that were design to target specific challenges in SED. We identified some challenges related to the localization of the sound events in time (working with longer clips, detection at different instants of the clip, detection of long sound events). Some of these challenge might also be emphasized by the metric itself as shown in Ferroni et al.~\cite{ferroni_2021} so this should be investigated further. We also analyzed the performance of SED systems with reverberated clips with non target sound events. Reverberation is consistently degrading the performance and some approaches as multi-condition training or dereverberation could be explored to solve this issue. SSep seems to help increasing the robustness of the SED systems to non-target sound events but its optimal integration with SED still needs to be investigated further.
\bibliographystyle{IEEEbib}
\bibliography{refs}

\begin{thebibliography}{10}

\bibitem{virtanen2018computational}
T.~Virtanen, M.~D. Plumbley, and D.~Ellis,
\newblock {\em Computational analysis of sound scenes and events},
\newblock Springer, 2018.

\bibitem{Bello:SONYC:CACM:18}
J.~P. Bello, C.~Silva, O.~Nov, R.~L. DuBois, A.~Arora, J.~Salamon, C.~Mydlarz,
  and H.~Doraiswamy,
\newblock ``{SONYC}: A system for the monitoring, analysis and mitigation of
  urban noise pollution,''
\newblock {\em Communications of the ACM}, In press, 2018.

\bibitem{radhakrishnan2005audio}
R.~Radhakrishnan, A.~Divakaran, and A.~Smaragdis,
\newblock ``Audio analysis for surveillance applications,''
\newblock in {\em Proc. WASPAA}. IEEE, 2005, pp. 158--161.

\bibitem{serizel2016}
R.~Serizel, V.~Bisot, S.~Essid, and G.~Richard,
\newblock ``{Machine listening techniques as a complement to video image
  analysis in forensics},''
\newblock in {\em Proc. ICIP}, 2016, pp. 948--952.

\bibitem{zigel2009method}
Y.~Zigel, D.~Litvak, and I.~Gannot,
\newblock ``A method for automatic fall detection of elderly people using floor
  vibrations and sound—proof of concept on human mimicking doll falls,''
\newblock {\em IEEE Transactions on Biomedical Engineering}, vol. 56, no. 12,
  pp. 2858--2867, 2009.

\bibitem{serizel2018_DCASE}
R.~Serizel, N.~Turpault, H.~Eghbal-Zadeh, and A.~Parag~Shah,
\newblock ``{Large-Scale Weakly Labeled Semi-Supervised Sound Event Detection
  in Domestic Environments},''
\newblock in {\em Proc. DCASE Workshop}, 2018.

\bibitem{Turpault2019_DCASE}
N.~Turpault, R.~Serizel, A.~Parag~Shah, and J.~Salamon,
\newblock ``{Sound event detection in domestic environments with weakly labeled
  data and soundscape synthesis},''
\newblock in {\em {Proc. DCASE Workshop}}, 2019.

\bibitem{turpault:hal-02891700}
N.~Turpault, S.~Wisdom, H.~Erdogan, J.~R. Hershey, R.~Serizel, E.~Fonseca,
  P.~Seetharaman, and J.~Salamon,
\newblock ``{Improving Sound Event Detection In Domestic Environments Using
  Sound Separation},''
\newblock in {\em Proc. DCASE workshop}, 2020.

\bibitem{salamon2017scaper}
J.~Salamon, D.~MacConnell, M.~Cartwright, P.~Li, and J.~P. Bello,
\newblock ``Scaper: A library for soundscape synthesis and augmentation,''
\newblock in {\em Proc. WASPAA}. IEEE, 2017, pp. 344--348.

\bibitem{Serizel2020_ICASSP}
R.~Serizel, N.~Turpault, A.~Shah, and J.~Salamon,
\newblock ``{Sound event detection in synthetic domestic environments},''
\newblock in {\em {Proc. ICASSP}}, 2020.

\bibitem{mesaros_metrics_2016}
A.~Mesaros, T.~Heittola, and T.~Virtanen,
\newblock ``Metrics for polyphonic sound event detection,''
\newblock {\em Applied Sciences}, vol. 6, no. 6, pp. 162, 2016.

\bibitem{ferroni_2021}
G.~Ferroni, N.~Turpault, J.~Azcarreta, F.~Tuveri, \c{C}a\u{g}da\c{s} Bilen,
  R.~Serizel, and S.~Krstulovi\'{c},
\newblock ``Improving sound event detection metrics: insights from dcase
  2020,''
\newblock submitted to ICASSP 2021.

\bibitem{turpault:hal-02160855}
N.~Turpault, R.~Serizel, A.~Parag~Shah, and J.~Salamon,
\newblock ``{Sound event detection in domestic environments with weakly labeled
  data and soundscape synthesis},''
\newblock in {\em Proc. DCASE Workshop}, 2019.

\bibitem{Gemmeke2017audioset}
J.~F. Gemmeke, D.~P.~W. Ellis, D.~Freedman, A.~Jansen, W.~Lawrence, R.~C.
  Moore, M.~Plakal, and M.~Ritter,
\newblock ``Audio set: An ontology and human-labeled dataset for audio
  events,''
\newblock in {\em Proc. ICASSP}, 2017.

\bibitem{font2013freesound}
F.~Font, G.~Roma, and X.~Serra,
\newblock ``Freesound technical demo,''
\newblock in {\em Proc. ACMM}. ACM, 2013, pp. 411--412.

\bibitem{fonseca2020fsd50k}
E.~Fonseca, X.~Favory, J.~Pons, F.~Font, and X.~Serra,
\newblock ``{FSD50K}: an open dataset of human-labeled sound events,''
\newblock in {\em arXiv:2010.00475}, 2020.

\bibitem{Dekkers2017}
G.~Dekkers, S.~Lauwereins, B.~Thoen, M.~W. Adhana, H.~Brouckxon, T.~van
  Waterschoot, B.~Vanrumste, M.~Verhelst, and P.~Karsmakers,
\newblock ``The {SINS} database for detection of daily activities in a home
  environment using an acoustic sensor network,''
\newblock in {\em Proc. DCASE Workshop}, 2017, pp. 32--36.

\bibitem{mesaros2016tut}
A.~Mesaros, T.~Heittola, and T.~Virtanen,
\newblock ``Tut database for acoustic scene classification and sound event
  detection,''
\newblock in {\em Proc. EUSIPCO}. IEEE, 2016, pp. 1128--1132.

\bibitem{wisdom2020fuss}
S.~Wisdom, H.~Erdogan, D.~P.~W. Ellis, R.~Serizel, N.~Turpault, E.~Fonseca,
  J.~Salamon, P.~Seetharaman, and J.~R. Hershey,
\newblock ``What's all the {FUSS} about free universal sound separation
  data?,''
\newblock {\em In preparation}, 2020.

\bibitem{kavalerov2019universal}
I.~Kavalerov, S.~Wisdom, H.~Erdogan, B.~Patton, K.~Wilson, J.~Le~Roux, and
  J.~R. Hershey,
\newblock ``Universal sound separation,''
\newblock in {\em Proc. WASPAA}, 2019.

\bibitem{Fonseca2017freesound}
E.~Fonseca, J.~Pons, X.~Favory, F.~Font, D.~Bogdanov, A.~Ferraro, S.~Oramas,
  A.~Porter, and X.~Serra,
\newblock ``Freesound datasets: a platform for the creation of open audio
  datasets,''
\newblock in {\em Proc. ISMIR}, 2017, pp. 486--493.

\bibitem{serizel2019}
R.~Serizel and N.~Turpault,
\newblock ``{Sound Event Detection from Partially Annotated Data: Trends and
  Challenges},''
\newblock in {\em {Proc. IcETRAN conference}}, 2019.

\bibitem{Miyazaki2020}
K.~Miyazaki, T.~Komatsu, T.~Hayashi, S.~Watanabe, T.~Toda, and K.~Takeda,
\newblock ``Convolution-augmented transformer for semi-supervised sound event
  detection,''
\newblock Tech. {R}ep., DCASE2020 Challenge, 2020.

\bibitem{Hao2020}
J.~Hao, Z.~Hou, and W.~Peng,
\newblock ``Cross-domain sound event detection: from synthesized audio to real
  audio,''
\newblock Tech. {R}ep., DCASE2020 Challenge, 2020.

\bibitem{Ebbers2020}
J.~Ebbers and R.~Haeb-Umbach,
\newblock ``Convolutional recurrent neural networks for weakly labeled
  semi-supervised sound event detection in domestic environments,''
\newblock Tech. {R}ep., DCASE2020 Challenge, 2020.

\bibitem{Koh2020}
C.-Y. Koh, Y.-S. Chen, S.-E. Li, Y.-W. Liu, J.-T. Chien, and M.~R. Bai,
\newblock ``Sound event detection by consistency training and pseudo-labeling
  with feature-pyramid convolutional recurrent neural networks,''
\newblock Tech. {R}ep., DCASE2020 Challenge, 2020.

\bibitem{Yao2020}
T.~Yao, C.~Shi, and H.~Li,
\newblock ``Sound event detection in domestic environments using dense
  recurrent neural network,''
\newblock Tech. {R}ep., DCASE2020 Challenge, 2020.

\bibitem{Chan2020}
T.~K. Chan, C.~S. Chin, and Y.~Li,
\newblock ``Semi-supervised nmf-cnn for sound event detection,''
\newblock Tech. {R}ep., DCASE2020 Challenge, 2020.

\bibitem{Liu2020}
Y.~Liu, C.~Chen, J.~Kuang, and P.~Zhang,
\newblock ``Semi-supervised sound event detection based on mean teacher with
  power pooling and data augmentation,''
\newblock Tech. {R}ep., DCASE2020 Challenge, 2020.

\bibitem{HouZ2020}
Z.~Hou, J.~Hao, and W.~Peng,
\newblock ``Author guidelines for dcase 2020 challenge technical report,''
\newblock Tech. {R}ep., DCASE2020 Challenge, 2020.

\bibitem{Huang2020}
Y.~Huang, L.~Lin, S.~Ma, X.~Wang, H.~Liu, Y.~Qian, M.~Liu, and K.~Ouch,
\newblock ``Guided multi-branch learning systems for dcase 2020 task 4,''
\newblock Tech. {R}ep., DCASE2020 Challenge, 2020.

\bibitem{Cornell2020}
S.~Cornell, G.~Pepe, E.~Principi, M.~Pariente, M.~Olvera, L.~Gabrielli, and
  S.~Squartini,
\newblock ``The univpm-inria systems for the dcase 2020 task 4,''
\newblock Tech. {R}ep., DCASE2020 Challenge, 2020.

\bibitem{turpault:hal-02891665}
N.~Turpault and R.~Serizel,
\newblock ``{Training Sound Event Detection On A Heterogeneous Dataset},''
\newblock in {\em Proc. {DCASE Workshop}}, 2020.

\bibitem{PSDS:2020}
C.~{Bilen}, G.~{Ferroni}, F.~{Tuveri}, J.~{Azcarreta}, and S.~{Krstulović},
\newblock ``A framework for the robust evaluation of sound event detection,''
\newblock in {\em Proc. ICASSP}, 2020, pp. 61--65.

\end{thebibliography}
\end{sloppy}
\end{document}